# Insights on Skyrme parameters from GW170817

C.Y. Tsang, M.B. Tsang[#], Pawel Danielewicz, and W.G. Lynch

*National Superconducting Cyclotron Laboratory and the Department of Physics and Astronomy Michigan State University, East Lansing, MI 48824 USA*

F.J. Fattoyev

*Physics Department, Manhattan College Riverdale, NY 10471, USA*

Abstract

The binary neutron-star merger event, GW170817, has cast a new light on nuclear physics research. Using a neutron-star model that includes a crust equation of state (EoS), we calculate the properties of a 1.4 solar-mass neutron star. The model incorporates more than 200 Skyrme energy density functionals, which describe nuclear matter properties, in the outer liquid core region of the neutron star. We find a power-law relation between the neutron-star tidal deformability, $\Lambda$, and the neutron-star radius, $R$. Without an explicit crust EoS, the model predicts smaller $R$ and the difference becomes significant for stars with large radii. To connect the neutron star properties with nuclear matter properties, we confront the predicted values for $\Lambda$, against the Taylor expansion coefficients of the Skyrme interactions. There is no pronounced correlation between Skyrme parameters in symmetric nuclear matter and neutron star properties. However, we find the strongest correlation between $\Lambda$ and $K_{sym}$, the curvature of the density dependence of the symmetry energy at saturation density. At twice the saturation density, our calculations show a strong correlation between $\Lambda$ and total pressure providing guidance to laboratory nucleus-nucleus collision experiments.

[#]Correspondence author: tsang@nscl.msu.edu

I. **Introduction**

Understanding the nature of dense nuclear matter and neutron stars is a compelling objective of nuclear science [1]. The recent observation of the neutron-star merger event, GW170817 provides important new constraints on the properties of neutron matter [2, 3]. Specifically, it provides a fertile ground for scientists to study the equation of state (EoS) of nuclear matter [2-11]. During the inspiral phase of a neutron-star merger, the gravitational field of each neutron star induces a tidal deformation in the other, changing the quadrupole moment of the system [2-4]. The tidal deformation of a neutron star is intimately related to the bulk properties of near pure neutron matter and its EoS. It also provides constraints on the neutron star radii [6-20]. In this paper, we will review how the new merger observable such as the tidal deformability correlates with nuclear physics parameters in nuclear EoS constructed from commonly used Skyrme interactions and how the insights gained can be used to guide nuclear physics experiments designed to constrain the symmetry energy terms of the nuclear EoS.

This paper is organized as follow: After the introduction, in Section II, we begin with a brief description of the neutron star model that is used to calculate the neutron star properties, such as tidal deformability and radii, using a large collection of Skyrme interactions. In Section III, we briefly describe the Skyrme models and the parameters from Taylor expansion of the symmetric matter and symmetry energy terms around saturation density. In the same section, we examine the correlation between the neutron-star properties and the Taylor expansion coefficients. In Section IV, the neutron star model is used to determine the density region that can best be studied in nucleus-nucleus collisions. Finally we summarize our work in Section V.

II. **Neutron-Star Model**

The influence of the EoS of neutron stars on the gravitational wave signal during inspiral is contained in the dimensionless tidal deformability, also known as tidal polarizability,

$$\Lambda = \frac{2}{3}k_2 \left(\frac{c^2 R}{GM}\right)^5, \qquad (1)$$

where $R$ and $M$ are the radius and mass of a neutron star, $G$ is the gravitational constant and $k_2$, is the dimensionless Love number [4] that also depends on $R$ [21]. The predicted mass-radius *(M-R)* relation and therefore $\Lambda$, is strongly correlated with the neutron-star matter EoS [22-25]. Recent analysis of the neutron star merger event, GW170817 extracted a range of possible Equations of State parameterized in terms of the dependence of pressure on density from sub-saturation to six times the saturation density [3].

To investigate which nuclear physics properties are tied to the tidal deformability, we incorporate the nuclear physics EoS in a neutron-star model code developed by Fattoyev [26] which solves the Tolman-Oppenheimer-Volkoff equation to calculate both neutron-star properties and nuclear physics observables. We choose the Skyrme density functionals [27] as the EoS in the liquid core of the neutron star. Below is a brief description of the model that assumes a neutron star consisting of an outer crust, an inner curst, an outer core and an inner core. Different forms of the EoS are employed in these different density regions.

1) For the outer crust, which consists of a Coulomb lattice of neutron-rich nuclei embedded in a degenerate electron gas, we use the well-known crust EoS prescribed in the seminal work of Baym, Pethick and Sutherland [28]. Here, the pressure of the electron gas provides a dominant contribution. As the density increases towards the boundary between the outer and the inner crust, the nuclei become progressively more neutron-rich until the neutron drip line is reached at the bottom of the outer crust.

2) The thickness of the inner crust constitutes about 5% to the thickness of the stellar crust; most of the inner crust is occupied by a lattice of spherical nuclei. Calculations predict that the bottom layers of the inner crust consist of complex and exotic structures that are collectively known as nuclear pasta [22, 29-32]. Whereas significant progress has been made in simulating this exotic region, the detailed EoS is still uncertain. In this region, the electron gas dominates the pressure; uncertainties in the nuclear contribution to the inner crust equation of state change the radius by less than 1% [33]. Considering the negligible impact that the nuclear EoS in this region has on the deformability and radii, we use the EoS from Negele et al. [34] of the form $P = A + K \left(\frac{E}{V}\right)^{\frac{4}{3}}$ [35, 36], corresponding to the EOS of relativistic Fermi gas, to connect the outer crust to the outer core. Here, E/V is the density of the total energy including mass. The parameters $A$ and $K$ are varied to match the crustal EoS to that of a liquid core at the crust-core transition density, $\rho_{TD}$, which occurs around $0.5\rho_0$ [37]. (An empirical formula is used to determine the crust-core transition density $\rho_{TD} = (-3.75 \times 10^{-4}L + 0.0963)$ fm$^{-3}$ where $L$, the slope of the symmetry energy at saturation density, is defined in the next section).

3) In the liquid core region of $\rho_{TD} < \rho < 3\rho_0$, which corresponds to the outer core, we use nuclear Skyrme density functionals. We require beta equilibrium between nuclear, electronic and muonic matter in this region, by minimizing energy density while maintaining charge conservation. Both electrons and muons are described by relativistic Fermi gases.

4) The density region above $3\rho_0$ mainly affects the maximum neutron-star mass and does not affect the relevant properties of the 1.4 solar-mass neutron stars considered in this work. The GW170817 has also stimulated a lot of work to constraint the maximum neutron-star mass with predictions in the range of 2.15-2.32 [38-44] solar mass. For the high density region of $\rho > 3\rho_0$, a polytropic EoS [24, 26] of the form $K'\rho^\gamma$ is used to extend the EoS to the central density region of a neutron star. The constants $K'$ and $\gamma$ are fixed by the conditions that the pressure at thrice the normal nuclear density, $P(3\rho_0)$, matches the pressure from the Skyrme density functionals and that the polytrope pressure at $7\rho_0$ is such that the EoS can support a 2.17 solar-mass neutron star [38]. This maximum mass is chosen as it is close to the heaviest neutron star observed [45, 46]. Our calculations show that the conclusion of the present work does not change if a different maximum mass is chosen. For example, if we do not include polytropic EoS, but rather just extend the Skyrme EoS beyond $3\rho_0$, with different maximum masses, the conclusions of this study do not change. This is because $\Lambda$ is not sensitive to the details of EoS at very high density regions. The fact that the prior and post distributions of pressure at high density regions obtained in Ref. [3] are not very different, is consistent with our results.

The EoS in different density regions are represented by different color curves in Fig. 1. At the lowest densities, the crust EoS are represented by gold color lines. The polytropic EoS that connect the crust to the outer core are represented by the green curves. The density region of the outer core, $0.5\rho_0$ to $3\rho_0$, is similar to that found in the nuclear matter environment. Here, we use the Skyrme interactions (blue curves) [27, 47, 48] to connect nontrivial nuclear physics observables to 1.4 solar-mass neutron-star observables. Above $3\rho_0$ the polytropic EoS are plotted in red. The Skyrme interactions that generate negative pressure at $3\rho_0$ or other densities would not support a 2.17 solar mass neutron star [38] and are excluded.

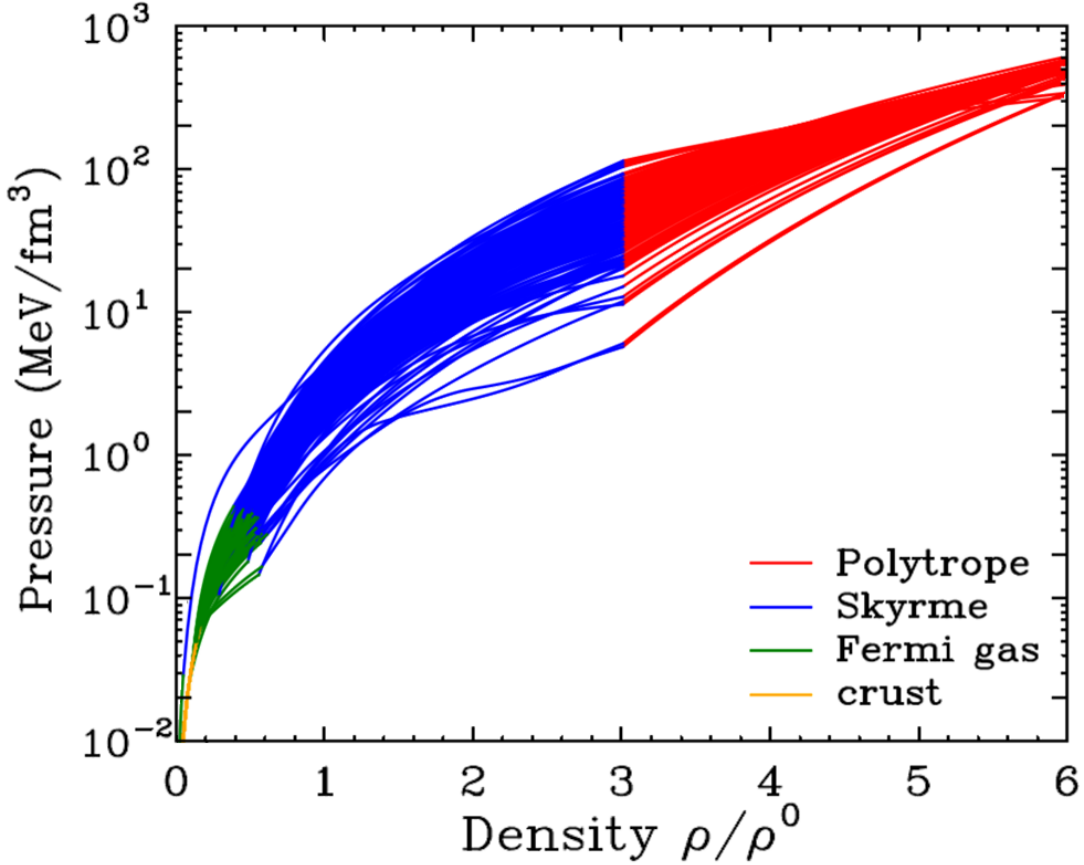

Figure 1: Equation of State used in different density regions in the neutron star, specifically the crust region (yellow), relativistic Fermi gas polytropic EoS (green), Skyrme EoS (blue) and high density polytropes (red). See text for details.

From the collection of 248 Skyrme interactions from Refs. [27, 47, 48], 182 of them can support a 2.17 solar mass neutron star. Each interaction, represented by an open circle in Fig. 2, gives rise to a unique prediction for the neutron-star radius and tidal deformability. The trend exhibited by the open circles reflects the fact that tidal deformability and neutron-star radius are correlated as described by Eq. (1). The solid line corresponding to $\Lambda = 5.9 \times 10^{-5} R^{6.26}$ is the best fit going through the calculated points. The value of the exponent for $R$ is not exactly 5 reflects the dependence of the Love number, $k_2$, on $R$ [21]. If we neglect the crust in our calculations, the blue dashed curve, $\Lambda = 2.9 \times 10^{-5} R^{6.63}$, represents the best fit. The larger power law exponent suggests a stiffer rise of $\Lambda$ with $R$. Thus, calculations without a crust, produce significantly smaller radii [10, 11] at high $\Lambda$ values.

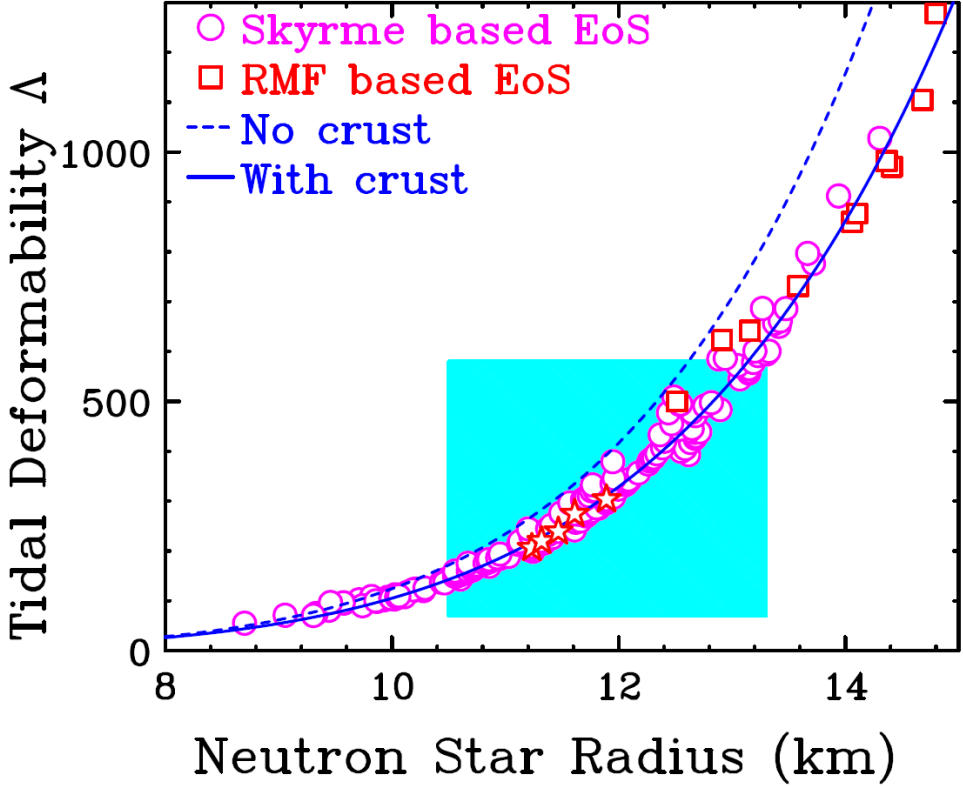

**Fig. 2:** Correlation between neutron-star tidal deformability and radius for 1.4 solar-mass neutron stars. Current calculations are represented by open circles and those from Ref. [6] by open squares. The light blue shaded area represents constraint from recent GW170817 analysis [3]. Five interactions, KDE0v1, LNS, NRAPR, SKRA, QMC700, deemed as the best in Ref. [27] in describing the properties of symmetric matter and calculated pure neutron matter, are represented with red stars. The solid curve is the best fit to our results with crust while the dashed curve is the best fit if no crust EoS is included in our neutron star model.

The increase in crustal thickness with neutron-star radius is consistent with Ref. [33], which shows that the crust thickness increases inversely with neutron-star compactness (M/R). The crust thickness contributes to the total radius, but does not affect the total mass and depends little on uncertainties in the crustal EoS, which is consistent with the findings of Ref. [49]. In the region of large tidal deformability, our results are consistent with those from EoS based on relativistic mean-field interactions [8] following analogous methodology and represented by the open red squares in Fig. 2. The range of the values $\Lambda=70-580$ and $R = 10.5-13.3$ $km$ obtained from the GW170817 analysis in Ref. [3] is represented by the large light blue-shaded square.

Our calculations lie nearly diagonally across the box. About 130 interactions are inside the GW constraint. At the intersection, our results suggest a lower limit of $\Lambda$ occurring around 100.

### III. Skyrme Interactions

One advantage of Skyrme nuclear density functionals is that many different Skyrme interactions have been developed to calculate nuclear properties and their use is well documented in the literature [27, 47, 48]. A detailed review of Skyrme models tested to be consistent with selected nuclear properties can be found in Ref. [27]. Eleven criteria that represent properties of symmetric nuclear matter and pure neutron matter are used to evaluate 240 Skyrme interactions. Five interactions, KDE0v1, LNS, NRAPR, SKRA, QMC700, which satisfy nearly all the 11 constraints, are highlighted as red stars in all the figures. Taken literally, they form a tight constraint on $\Lambda$ = 200-350 with the associated $R$ = 11-12 km.

There had been many efforts to characterize the variables in the Skyrme interactions in terms of parameters of infinite nuclear matter. Here we will introduce these parameters using similar convention as that used in Ref. [27]. Within the parabolic approximation [50], the EoS of cold nuclear matter, expressed as the energy per nucleon of the hadronic system, $\varepsilon(\rho, \delta)$, can be divided into a symmetric-matter contribution, $\varepsilon_{SNM}(\rho)=\varepsilon(\rho, \delta=0)$, that is independent of the neutron-proton asymmetry, and a symmetry-energy term, $\varepsilon_{sym}(\rho,\delta)=S(\rho)\delta^2$, proportional to the square of the asymmetry, $\delta=(\rho_n-\rho_p)/\rho$, following the expansion [51]:

$$\varepsilon(\rho, \delta) = \varepsilon(\rho, \delta=0) + S(\rho)\delta^2 + O(\delta^4) + ... \quad , \qquad (2)$$

where $\rho_n$, $\rho_p$ and $\rho=\rho_n+\rho_p$ are the neutron, proton and nucleon number densities, respectively, and $S(\rho)$ is the density dependence of the symmetry energy. Relative to $S(\rho)$, the contributions to the EoS from known higher order terms, $O(\delta^4)$ and above, are small for $\rho<\rho_0$, less than *15%* at $2\rho_0$ [50]. Employing the Taylor expansion around the saturation density, $\rho_0$, and defining $x=(\rho-\rho_0)/(3\rho_0)$, the density dependence of both $\varepsilon_{SNM}(\rho)$ and $S(\rho)$ can be written as:

$$\varepsilon_{SNM}(\rho)=E_0+\frac{1}{2}K_0 x^2+\frac{1}{6}Q_0 x^3+ O(x^4) \quad , \qquad (3)$$

$$S(\rho)=J+Lx+\frac{1}{2}K_{sym}x^2+\frac{1}{6}Q_{sym}x^3+O(x^4), \qquad (4)$$

where $K_0$ and $Q_0$ in Eq. (3) are known as the incompressibility and skewness coefficients of the symmetric nuclear matter while $J, L, K_{sym}, Q_{sym}$ in Eq. (4) are the symmetry energy, symmetry slope, symmetry curvature or incompressibility, and symmetry skewness coefficients, respectively, at saturation density. By taking advantage of the large range of Skyrme parameters used in this work, we can explore the correlations of the set ($K_{sym}$, $Q_{sym}$, $K_0$, $Q_0$, $J$, $L$) with the neutron star properties, and here, specifically the tidal deformability, $\Lambda$. ($\Lambda$ and $R$ are related in Eq. (1) so similar conclusions can be applied to $R$).

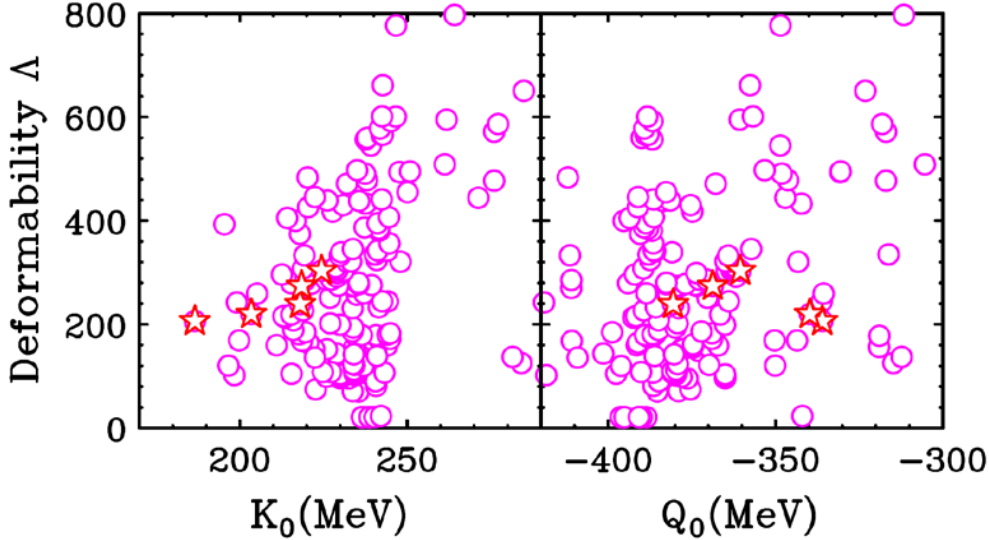

**Fig. 3** Correlation between the neutron star deformability $\Lambda$ for 1.4 solar-mass neutron stars and the incompressibility parameter $K_0$ (left panel) and skewness parameter $Q_0$ (right panel) defined in Eq. (3), for the symmetric matter EoS of Skyrme functionals used in the study. The symbols follow the same convention as in Fig. 2.

First we explore the connection to the symmetric nuclear matter parameters in $\varepsilon_{SNM}(\rho)$ in Eq. (3). Fig. 3 shows the plots of $\Lambda$ vs. $K_0$ (left panel) and $Q_0$ (right panel). The value of $K_0$ [27], also known as compressibility, has been fairly well determined experimentally to be *230±30 MeV*. Most of the Skyrme interactions studied here cluster around $K_0$ ~ *240 MeV* and, within this tight bound on $K_0$, show no strong correlation with $\Lambda$. Similarly, $Q_0$ cluster around 380 MeV and again show no strong correlation with $\Lambda$. The observation that the tidal deformability, $\Lambda$, is not strongly connected to the parameters, $K_0$ and $Q_0$, which characterize the symmetric nuclear matter, is consistent with conclusions of previous studies [5]. This implies that it would be difficult to extract properties of the symmetric nuclear matter from the tidal deformability even though the contribution of the symmetric matter EoS to pressure supporting the star can be comparable to that provided by the symmetry energy.

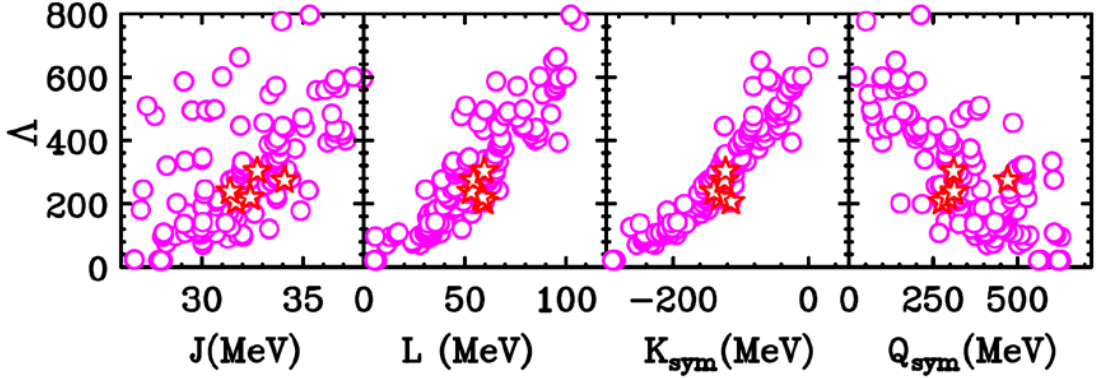

**Fig. 4:** The four panels show the correlation between the neutron star deformability $\Lambda$ for 1.4 solar-mass neutron stars and Taylor expansion coefficients (from left to right) $J$, $L$, $K_{sym}$ and $Q_{sym}$ defined in Eq. (4) obtained for the Skyrme functionals used in the study. The symbols follow the same convention as in Fig. 2.

Next we explore the importance of the parameters in symmetry energy term, $J$, $L$, $K_{sym}$, and $Q_{sym}$ in Fig. 4. The abscissa scales are chosen to represent the respective ranges of values found in Ref. [27] so that the correlations between the plots are comparable. The correlations between $\Lambda$ and symmetry energy parameters are much stronger than those for symmetric nuclear matter. The correlation is the strongest for $K_{sym}$. Since the second-order term, $K_{sym}$ impacts the higher densities more, it is not surprising that $K_{sym}$ should have stronger influence on $\Lambda$ than the slope $L$, which is the first derivative and governs the pressure in the vicinity of saturation density. The much weaker sensitivity to $Q_{sym}$ probably reflects the fact that the range of $Q_{sym}$ is not well determined and/or that its impact on the pressure above $3\rho_0$ is small. This is consistent with our observation in Section II that the EoS above $3\rho_0$ does not affect the tidal deformability. For simplicity, most effective interactions have a limited number of terms; thus the various coefficients of their Taylor expansion are often correlated [52]. Different models may have different correlations between $J$, $L$, $K_{sym}$, $Q_{sym}$. For example, a stronger correlation between $\Lambda$ and $L$ than $\Lambda$ and $K_{sym}$ is reported in Ref. [5, 53]. Additional analyses with different forms of the nuclear effective interactions, especially those created assuming minimal correlation between these coefficients may provide guidance on the biases that may be introduced by specific choices of nuclear interactions.

### IV. Laboratory Observable sensitive to the tidal deformability

Analysis of GW170817 data has provided new constraints on the overall pressure as a function of density [3]. The density range in the GW constraint, from 0.5 to 6 times the saturation density, $\rho_0$, is rather wide. In this section, we search for the best condition for laboratory observable to provide constraints relevant to the tidal deformability.

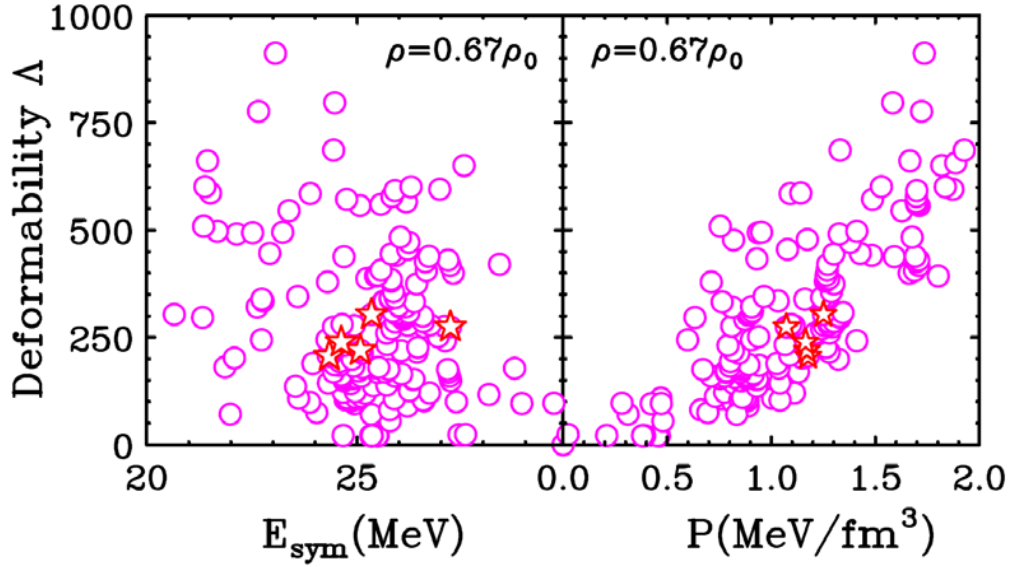

**Fig. 5**: Neutron-star tidal deformability vs. symmetry energy (left panel) and vs. pressure (right panel) at $0.67\rho_0$ for 1.4 solar-mass neutron stars. The symbols follow the same convention as in Fig. 2.

The energy of neutron matter is experimentally best constrained at $0.67\rho_0$ [47, 54] where an accurate value (~25 MeV) of the symmetry energy has been derived from the analysis of nuclear masses [47, 55] and isobaric analog states [48, 52]. However, there is no correlation between $\Lambda$ and the symmetry energy at this density as shown in the left panel of Fig. 5. The correlation is slightly stronger between the pressure and the deformability as shown in the right panel. The pressure of 0.56 - 1.12 MeV/fm3 at $0.67\rho_0$ obtained in Ref. [3] suggests $\Lambda$ is >100 even though the spread in $\Lambda$ is large. This lower bound is consistent with the observation in Fig. 2. Currently there are no laboratory constraints on the total pressure at subsaturation density. Due to the small pressure value, it may be very difficult to measure.

At supra-saturation density, pressure at $2\rho_0$ has been identified to be sensitive to the neutron-star radius [10, 11, 56]. At this density our calculations (open symbols in Fig. 6) show some correlations between $\Lambda$ and the symmetry energy, but much stronger correlation is found between $\Lambda$ and the total pressure of each Skyrme functional. This strong correlation combined with $\Lambda$-$R$ correlation in Fig. 2 implies that a strong correlation exists between the neutron star radius and the neutron star pressure at $2\rho_0$, as demonstrated previously in Ref. [56]. Past experiments that measure transverse and elliptical flows from Au+Au collisions have extracted pressure vs. density constraints for symmetric matter [57]. Similar methodology can be used to extract the symmetry pressure. Experimental efforts are already underway to extract the EoS of asymmetric matter at this density region [58-61].

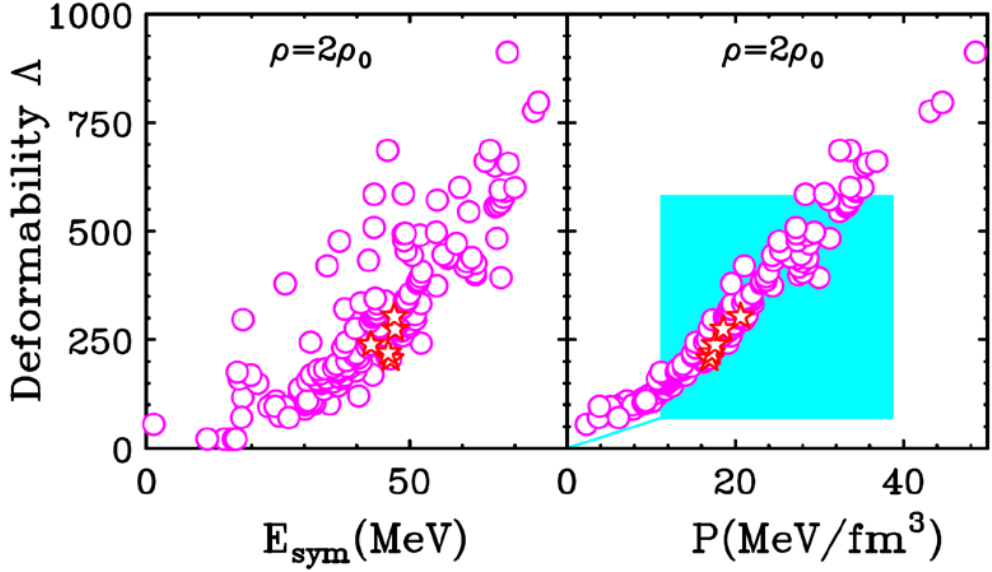

Fig. 6: Neutron-star tidal deformability vs. symmetry energy (left panel) and vs. pressure (right panel) at $2\rho_0$ for 1.4 solar-mass neutron stars.

### V. Summary

In summary, we have used a neutron-star model that employs nuclear equation of state in the outer-core region of the neutron-star to correlate both neutron-star and nuclear properties. Our calculations also show that the crust EoS affects the neutron star radii, and that the pressure at density above $3\rho_0$ has stronger impact on the maximum mass of a neutron star than on the tidal deformability of 1.4 solar mass neutron star.

Consistent with previous work, understanding the origin of neutron star tidal deformability requires a better understanding of the density dependence of the symmetry energy. For the symmetry energy term, we have explored the correlations between the tidal deformability and the first, second and third derivative coefficients in the Taylor expansion of the symmetry energy. For Skyrme interactions, both $L$ and $K_{sym}$ are strongly correlated with the tidal deformability $\Lambda$. The correlation is stronger for $K_{sym}$, as it more strongly influences higher density regions. However, the predicted sensitivity may not be the same for other effective interactions. Analyses with different forms of the nuclear effective interactions can provide guidance on the biases that may be introduced by specific choices of nuclear interactions. Future work should explore equations of state that have minimal correlations between these coefficients.

Finally, while low density constraints on symmetry energy extracted from data do not correlate strongly with neutron-star properties, heavy ion collision experiments testing twice the normal nuclear matter density may provide tighter constraints on the tidal deformability and thus the corresponding neutron-star radii.

**Acknowledgement**:

This work is supported by the US National Science Foundation under Grant No. PHY-1565546, and U.S. Department of Energy (Office of Science) under Grant Nos. DE-SC0014530, DE-NA0002923, DE-SC0010209, and through travel support from CUSTIPEN (China-US Theory Institute for Physics with Exotic Nuclei) under the US Department of Energy Grant No. DE-FG02-13ER42025. This work was stimulated by discussions with participants at the INT-JINA Symposium "First multi-messenger observations of a neutron-star merger and its implications for nuclear physics".

**REFERENCES**

[1] J. Carlson et al. Prog. Part. Nucl. Phys. **94** (2017 ) 68.
https://science.energy.gov/~/media/np/nsac/pdf/2015LRP/2015_LRPNS_091815.pdf
Crossref

[2] B. P. Abbott et al. (LIGO+Virgo Collaborations) Phys. Rev. Lett. **119** (2017) 161101.
Crossref

[3] B. P. Abbott et al. (LIGO+Virgo Collaborations) Phys. Rev. Lett. **121** (2018) 161101.
Crossref

[4] K. Chatziioannou, C. J. Haster, and A. Zimmerman Phys. Rev. **D97** (2018) 104036.
Crossref

[5] N. B. Zhang, B. A. Li, and J. Xu Astrophys. J. **859** (2018) 90.
Crossref

[6] F. J. Fattoyev, J. Piekarewicz, and C. J. Horowitz Phys. Rev. Lett. **120** (2018) 172702.
Crossref

[7] E. Annala, T. Gorda, A. Kurkela, and A. Vuorinen Phys. Rev. Lett. (2018) **120** 172703.
Crossref

[8] E. R. Most, L. R. Weih, L. Rezzolla, and J. Schaffner-Bielich Phys. Rev. Lett. **120** (2018) 261103.
Crossref

[9] S. De, D. Finstad, J. M. Lattimer, D. A. Brown, E. Berger, and C. M. Biwer Phys. Rev. Lett. **121** (2018) 091102.
Crossref

[10] Y. Lim and J. W. Holt Phys. Rev. Lett. **121** (2018) 062701.
Crossref

[11] I. Tews, J. Margueron and S. Reddy Phys. Rev. C **98** (2018) 045804.
Crossref

[12] F.J. Fattoyev, W.G. Newton and B.A. Li Euro Phys. J. **A50** (2014) 45
Crossref

[13] R. Nandi, P. Char and S. Pal arXiv: 1809.07108
Crossref

[14] O. Lourenco, M. Dutra, C.H. Lenzi, S.K. Biswa, M. Bhuyan and D.P. Menezes arXiv: 1901.04529
Crossref

[15] C. Raithel, F. Ozel and D. Psaltis Astrophys. J. Lett. **857** (2018) L23
Crossref

[16] T. Malik, N. Alam, M. Fortin, C. Providencia, B.K. Agrawal, T.K. Jha, B. Kumar and S.K. Patra Phys. Rev. **C98** (2018) 035804
Crossref

[17] Y. Zhou, L.W. Chen and Z. Zhang arXiv: 1901.11364
Crossref

[18] D. Radice and L. Dai L. Eur. Phys. J. **A55** (2019) 50
Crossref

[19] A. Bauswein, O. Just, H. Janka and N. Stergioulas Astrophys. J. Lett. **850** (2017) L34
Crossref

[20] H. Tong, P.W. Zhao and J. Meng arXiv: 1903.05938v1
Crossref

[21] T. Hinderer, B. D. Lackey, R. N. Lang and J.S. Read Phys. Rev. **D81** (2010) 123016.
Crossref

[22] F. J. Fattoyev, C. J. Horowitz and B. Schuetrumpf Phys. Rev. **C95** (2017) 055804.
Crossref

[23] S. Postnikov, M. Prakash and J. M. Lattimer Phys. Rev. D (2010) **82** 024016.
Crossref

[24] J. M. Lattimer and M. Prakash Phys. Rep. **621** (2016) 127.
Crossref

[25] L. Lindblom Astrophys. J. **398** (1992) 569.
Crossref

[26] F. J. Fattoyev, J. Carvajal, W. G. Newton and B. A. Li Phys. Rev. **C87** (2013) 015806.
Crossref

[27] O. Dutra M, Lourenco, J. S. SaMartins, A. Delfino, J. R. Stone and P. D. Stevenson Phys. Rev. **C85** (2012) 035201.
Crossref

[28] G. Baym, C. Pethick and P. Sutherland Astrophys J. **170** 1971 299.
Crossref

[29] C. P. Lorenz, D. G. Ravenhall and C. J. Pethick Phys. Rev. Lett. **70** (1993) 379.
Crossref

[30] A. S. Schneider, C. J. Horowitz, J. Hughto and D. K. Berry Phys. Rev. C **88** (2013) 065807.
Crossref

[31] A. S. Schneider, D. K. Berry, C. M. Briggs, M. E. Caplan and C. J. Horowitz Phys. Rev. C **90** (2014) 055805.
Crossref

[32] D. K. Berry, M. E. Caplan, C. J. Horowitz, G. Huber and A. S. Schneider Phys. Rev. **C94** (2016) 055801.
Crossref

[33] H. Sotani H, K. Iida and K. Oyamatsu MNRAS **470** (2017) 4397.
Crossref

[34] J. W. Negele and D. Vautherin Nucl. Phys. **A207** (1973) 298.
Crossref

[35] B. Link, R. I. Epstein and J. M. Lattimer Phys. Rev. Lett. **83** (1999) 3362.
Crossref

[36] J. Carriere, C. J. Horowitz and J. Peikarewicz Astrophys. J. **593** (2003) 464.
Crossref

[37] C. Ducoin, J. Marugueron, C. Providência and I. Vidaña Phys. Rev. **C83** (2011) 045810.
Crossref

[38] B. Margalit and B. D. Metzger Astrophys. J. Lett. **850** (2017) L19.
Crossref

[39] Gordon Baym, Shun Furusawa, Tetsuo Hatsuda, Toru Kojo, Hajime Togashi, arXiv:1903.08963 (2019)
Crossref

[40] N. B. Zhang and B. A. Li, Euro Phys. J. A 55:39 (2019)
Crossref

[41] M. Shibata, S. Fujibayashi, K. Hotokezaka, K. Kiuchi, K. Kyutoku, Y. Sekiguchi and M. Tanaka, Phys. Rev. D 96, 123012 (2017)
Crossref

[42] L. Rezzolla, E.R. Most and L.R. Weih, Astrophys. J. 852, L25 (2018)
Crossref

[43] M. Ruiz and S.L. Shapiro, A. Tsokaros, Phys. Rev. D, 97, 021501(R) (2018)
Crossref

[44] E. P. Zhou and X. Zhou, A. Li, Phys. Rev. D 97, 083015 (2018)
Crossref

[45] P. B. Demorest, T. Pennucci, S. M. Ransom, M. S. E. Roberts, and J. W. T. Hessels, Nature 467, 1081 (2010).
Crossref

[46] J. Antoniadis, P. C. Freire, N. Wex, T. M. Tauris, R. S. Lynch, et al., Science 340, 6131 (2013)
Crossref

[47] B. A. Brown Phys. Rev. Lett. **111** (2013) 232502.
Crossref

[48] P. Danielewicz, P. Singh and J. Lee Nucl. Phys. **A958** (2017) 147.
Crossref

[49] I. Tews, J. Margueron and S. Reddy arXiv:1901.09874 (2019).
Crossref

[50] C. Gonzalez-Boquera, M. Centelles, X. Viñas and A. Rios Phys. Rev. **C96** (2017) 065806.
Crossref

[51] C. J. Horowitz et al. J. Phys. G: Nucl. Part. Phys. **41** (2014) 093001.
Crossref

[52] P. Danielewicz and J. Lee Nucl. Phys. **A922** (2014)1.
Crossref

[53] N. B. Zhang and B. A. Li J. Phys. G: Nucl. Part. Phys. **46** (2019) 014002.
Crossref

[54] W. G. Lynch and M. B. Tsang arXiv:1805.10757 (2018).
Crossref

[55] M. Kortelainen et al. Phys. Rev. **C82** (2010) 024313.
Crossref

[56] J. M. Lattimer and M. Prakash Astrophys. J. **550** (2001) 426.
Crossref

[57] Pawel Danielewicz, Roy Lacey, and William G. Lynch, Science 298, 1592 (2002).
Crossref

[58] M. B. Tsang et al. Phys. Rev. C **95** (2017) 044614.
Crossref

[59] R. Shane et al. Nucl. Instrum. Methods A **784** (2015) 513.
Crossref

[60] D. M. Cozma, Eur. Phys. J. **A54** (2018) 40.
Crossref

[61] P. Russotto et al. Phys. Rev. C **94** (2016) 034608.
Crossref